\documentclass[aps,pra,twocolumn,groupedaddress,showpacs]{revtex4}

\usepackage{graphicx}
\usepackage{amsmath}

\begin{document}


\title{Vortex sheet in rotating two-component Bose-Einstein condensates}


\author{Kenichi Kasamatsu$^{1}$}
\author{Makoto Tsubota$^{2}$}
\affiliation{$^1$Department of Physics, Kinki University, Higashi-Osaka, 577-8502, Japan \\
$^2$Department of Physics, Osaka City University, Sumiyoshi-Ku, Osaka 558-8585, Japan}


\date{\today}

\begin{abstract}
We investigate vortex states of immiscible two-component Bose-Einstein condensates 
under rotation through numerical simulations of the coupled Gross-Pitaevskii equations. 
For strong intercomponent repulsion, the two components undergo phase separation 
to form several density domains of the same component. In the presence of 
the rotation, the nucleated vortices are aligned between the domains to make 
up winding chains of singly quantized vortices, a {\it vortex sheet}, 
instead of periodic vortex lattices. The vortices of one component are located 
at the region of the density domains of the other component, which results in 
the serpentine domain structure. The sheet configuration is stable 
as long as the imbalance of the {\it intracomponent} parameter is small. 
We employ a planar sheet model to estimate the distance between 
neighboring sheets, determined by the competition between the surface 
tension of the domain wall and the kinetic energy of the superflow via quantized vortices. 
By comparing the several length scales in this system, the phase diagram of the 
vortex state is obtained.
\end{abstract}


\maketitle

\section{INTRODUCTION}
Diverse interests have been focused on the quantized vortices in trapped atomic 
Bose-Eistein condensates (BECs) \cite{kasareview}. This pristine Bose-condensed 
system is highly controllable, having a great advantage to investigate the 
physics of topological defects such as vortices. Rotation of the trapping potential can 
create a lattice of quantized vortices \cite{Madison}, whose properties are 
observable by direct imaging. 
The next step of this kind of study will be elucidation of the vortex states in 
multicomponent BECs \cite{Kasarev2}. 
A multicomponent order parameter allows the existence of the various unconventional 
topological defects, which have been discussed in the other condensed matter 
systems and cosmological theory \cite{Volovik,supercond}. 

Two-component BECs, the simplest example of the multicomponent condensates, have also 
attracted much interest to study the novel phenomena not found in the single component system. 
The typical example is phase separation of the binary BEC mixtures, where each component 
separates spatially due to the strong intercomponent repulsion. Several experiments have 
observed this phase separation \cite{Hall,Miesner,Modugno,Mertes,Thalhammer,Papp}, 
supported by the numerous theoretical studies \cite{Ho,Ersy,Pu,Tim,Ao,Trippenbach,Svidzinsky,Kasamatsu2}. 
We can now control miscibility or immiscibility of the two-component BEC, 
because intercomponent interaction is tunable by changing the $s$-wave scattering length 
via a Feshbach resonance, as demonstrated recently \cite{Thalhammer,Papp}.

Rotating two-component BECs could have a rich variety of vortex phases in addition to the conventional 
triangular lattice \cite{Mueller,Kasamatsu,Keceli,Woo,Bargi,Barnett}. Schweikhard {\it et al.} 
observed the transient structural transition of the vortex lattice in two-component 
BECs \cite{Schweikhard}. The richness of the vortex structure originates from the intercomponent 
interaction. For instance, with increasing the strength of intercomponent interaction from zero, 
the vortex lattice deforms from triangular to square, eventually evolving multiple vortex sheets 
with each component made up of chains of singly quantized vortices \cite{Mueller,Kasamatsu}.  
Therefore it is challenging to see experimentally these structural transitions by using 
two-component BECs with tunable interactions.

In this paper, we study the detail of the vortex sheet structure 
in rotating immiscible two-component BECs, 
which has not been explored so much. This system has three 
interatomic coupling constants denoted by 
$g_{1}$ and $g_{2}$ (for intracomponents), and $g_{12}$ (for intercomponent). 
We confine ourselves to the phase-separated regime; in a homogeneous system, 
the condition 
is given by \cite{Tim,Ao}
\begin{equation}
g_{12}^2 > g_{1} g_{2}.
\end{equation}
To give a clear picture of the vortex state, we first address the simulation results of the 
coupled Gross-Pitaevskii (GP) equations, discussing various configurations of the vortex 
states in immiscible BECs. Next, we give an analytical discussion of the sheet structure. 
The distance of the vortex sheet can be estimated by the ratio of the surface tension of the 
domain boundary and the kinetic energy of the superflow due to the vortices \cite{Parts}. 
Comparing the several length scales characteristic of our system, we can obtain the 
phase diagram of the vortex states, which provides observable conditions 
of the several vortex phases, such as multiple vortex sheets with stripe pattern, 
or ``serpentine" sheets, or the ``rotating droplets." 

There are other condensed-matter systems characterized by the 
multicomponent order parameters in which a vortex sheet is observable 
\cite{Parts,Matsunaga}. In Sec. \ref{helium3}, we give a brief 
review of a vortex sheet in superfluid $^{3}$He to help the following 
discussion. Section \ref{2BECcase} consists of three parts; after formulating 
our problem in Sec. \ref{formulation}, we present the numerical results 
in Sec. \ref{numerical} and analytical discussion in Sec. \ref{analytical} for 
the prolem of two-component BECs. 
We conclude this paper in Sec. \ref{concle}. 

\section{VORTEX SHEET IN A ROTATING SUPERFLUID HELIUM}\label{helium3}
\begin{figure}
\begin{center}
\includegraphics[width=0.8\linewidth,keepaspectratio]{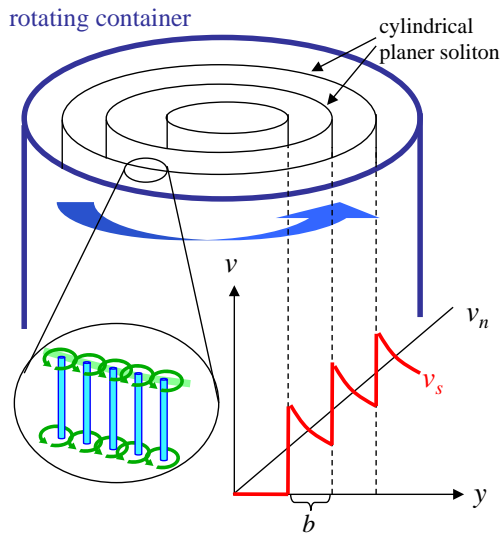}
\end{center}
\caption{(Color online) Vortex sheet array suggested by Landau and Lifshitz. 
Straight vortex lines are aligned along cylindrical planar soliton.
The schematic view of the radial velocity profile is also shown.} 
\label{sheetsche}
\end{figure}

The vortex sheet is well known from classical turbulence as a thin interface across which 
the tangential component of the flow velocity is discontinuous. 
In the limit of vanishing width of the sheet, the vorticity approaches infinity within the sheet.
Historically in superfluids, an array of vortex sheets was suggested as the possible 
ground state of rotating superfluid $^4$He \cite{Landau}. 
The schematic picture of the vortex sheet, suggested by Landau and Lifshitz, is shown in Fig. \ref{sheetsche}.
A vortex sheet consists of a chain of quantized vortices with the same orientation of circulation, 
which are confined by the cylindrical planar soliton. The tangential component of 
the superfluid velocity ${\bf v_s}$ discontinuously jumps across the sheet. 
It soon turned out, however, that in superfluid $^{4}$He a vortex sheet is unstable 
towards breaking up into separated vortex lines, namely forming a vortex lattice. 

In contrast, in superfluid $^3$He-A, the vortex sheet is stable due to 
the confinement of the vorticity within the topologically stable solitons \cite{Parts}. 
In this system, vortices are confined by the planar soliton sheets, 
which separate regions with opposite orientations of the orbital angular momentum 
($\hat{\bf l}$ vectors) of the Cooper pairs. In equilibrium, the vortex sheets uniformly 
fill the rotating container, forming accurately equidistant layers with the distance $b$.
Here, the calculation by Landau and Lifshitz is useful to explain 
the equilibrium distance between the planes of vortex sheet in $^3$He-A in a container 
rotating with $\Omega \hat{\bf z}$, even though their calculation did not take into 
account the quantization condition \cite{Landau}. 
The equilibrium distance $b$ is determined by the competition of the surface tension 
$\sigma$ of the soliton and the kinetic energy of the counterflow ${\bf v_n} - {\bf v_s}$ 
outside the sheet, where ${\bf v_n} $ is the normal fluid velocity. 
The planar sheet is assumed to be parallel to the $\hat{\bf x}$ axis for a large system.
The motion of the normal component makes a rigid body rotation 
$\nabla \times {\bf v_n} = 2 \Omega \hat{\bf z}$ and can be represented 
by the shear flow  ${\bf v_n} = - 2 \Omega y \hat{\bf x}$ parallel to the planes. 
In the region between nearest planes, the vortex-free superfluid velocity ${\bf v_s}$
is constant and equals the average ${\bf v_n}$ to minimize the counterflow. Thus the 
velocity jump across the sheet is $\Delta v_s = 2 \Omega b$. 
The counterflow energy per volume is 
$(1/b) \int \frac{1}{2} \rho_s ({\bf v_n} - {\bf v_s})^2 dy = \rho_s \Omega^2 b^2 / 6$, 
where $\rho_s = m n_s$ is the superfluid mass density. 
The surface energy per volume equals $\sigma/b$. Minimizing the sum of two 
contributions with respect to $b$, one obtains $b = (3 \sigma / \rho_s \Omega^2)^{1/3}$. 
For $\Omega = 1$ rad/s, this gives 0.3-0.4 mm, which was quantitatively confirmed 
by the NMR absorption measurement in superfluid $^3$He-A \cite{Parts}. 
The possible sheet configurations and their response for rotating drive were 
studied by numerical simulations in Ref. \cite{Eltsov}. 

\section{VORTEX SHEET IN TRAPPED TWO-COMPONENT BECS} \label{2BECcase}

\subsection{Coupled Gross-Pitaevskii equations} \label{formulation}
We consider atomic two-component BECs, characterized by the condensate 
wave functions $\Psi_{i}$ ($i=1,2$). The condensates are assumed to be confined 
by trapping potentials $V_{i}({\bf r})=m_{i}(\omega_{i}^{2} r^{2} 
+ \omega_{z i}^{2} z^{2})/2$; 
we neglect the shift of the potential minima for two components due to the 
gravitational effect or the difference of the hyperfine sublevels.
The potential is assumed to rotate at a rotation frequency $\Omega$ about the $z$ axis 
as ${\bf \Omega} = \Omega \hat{\bf z}$. 
Viewed from the frame of reference corotating with the trap potential, 
the GP energy functional of our problem reads
\begin{eqnarray}
E[\Psi_{1},\Psi_{2}] = \int d^{3} r \biggl[ \sum_{i=1,2} \Psi_{i}^{\ast} \biggl( - \frac{\hbar^{2} 
\nabla^{2}}{2m_{i}} + V_{i}  -\Omega L_{z} \nonumber \\ 
+ \frac{g_{i}}{2} |\Psi_{i}|^{2} \biggr) \Psi_{i} + g_{12} |\Psi_{1}|^{2} |\Psi_{2}|^{2} \biggr].   
\label{energyfunctio2}
\end{eqnarray}
Here, $m_{i}$ is a mass of the $i$th atom, 
$\Omega L_{z} = i \hbar \Omega (y \partial_{x} -x \partial_{y})$ is a centrifugal term, 
and the coefficients $g_{i}$ ($i=1,2$) and $g_{12}$ represent the atom-atom interactions. 
These interactions are expressed in terms of the s-wave scattering lengths $a_{1}$ and $a_{2}$ 
between atoms in the same component and $a_{12}$ between atoms in different component as 
\begin{equation}
g_{i} = \frac{4 \pi \hbar^{2} a_{i}}{m_{i}},  \hspace{5mm}
g_{12} = \frac{2 \pi \hbar^{2} a_{12}}{m_{12}},
\end{equation}
where $m_{12}^{-1}=m_{1}^{-1}+m_{2}^{-1}$ is the reduced mass. Each component interacts 
with the other through the intercomponent mean-field coupling $g_{12} |\Psi_1|^2 |\Psi_2|^2$, 
which yields structures and dynamics not found in a single-component BEC 
\cite{Hall,Miesner,Modugno,Mertes,Thalhammer,Papp,Ho,Ersy,Pu,Tim,Ao,
Trippenbach,Svidzinsky,Kasamatsu2}.

The properties of two-component BECs are described by the coupled GP equations, 
obtained from Eq. (\ref{energyfunctio2}) by using a variational procedure 
$i \hbar \partial \Psi_{i} = \delta E / \delta \Psi_{i}^{\ast}$ as 
\begin{eqnarray}
i \hbar \frac{\partial \Psi_{i}}{\partial t} = \biggl( -\frac{\hbar^{2} \nabla^{2}}{2m_{i}} 
+ V_{i} + g_{i}|\Psi_{i}|^{2} + g_{12}|\Psi_{3-i}|^{2} \nonumber \\
-\Omega L_{z} \biggr) \Psi_{i}. \label{bingp1td}
\label{bingp2td}
\end{eqnarray}
Since we are interested in the stationary solutions of this equation, we consider 
the time-independent coupled GP equations by substituting 
$\Psi_{i} ({\bf r},t) = \Psi_{i}({\bf r}) e^{-i \mu_{i} t / \hbar}$ as
\begin{eqnarray}
\mu_{i} \Psi_{i} = \biggl( -\frac{\hbar^{2} \nabla^{2}}{2m_{i}} + V_{i} 
+ g_{i}|\Psi_{i}|^{2} +g_{12}| \Psi_{3-i}|^{2} \nonumber \\
 -\Omega L_{z} \biggr) \Psi_{i} ,  \label{bingp1}
\end{eqnarray}
where we have introduced the Lagrange multiplier $\mu_{i}$ which represents 
the chemical potential and is determined so as to satisfy the conservation of particle number 
$N_{i} = \int d^3 r n_i = \int d^{3} r |\Psi_{i}|^{2}$. 

Even for $\Omega = 0$ the equilibrium solutions of the coupled GP equations (\ref{bingp1}) exhibit a 
rich variety of structures, depending on the various parameters 
of the system \cite{Ho,Ersy,Pu,Tim,Ao,Trippenbach,Svidzinsky}. 
In particular, the intercomponent interaction $g_{12}$ plays an important role 
in determining the ground state structure; when the intercomponent 
interaction is strongly repulsive, the two components phase separate. 
Adapting the Thomas-Fermi approximation, one can see that 
there is no solution with overlapping density profile when $g_{12}^2 > g_{1} g_{2}$ 
\cite{Tim,Ao,Trippenbach,Svidzinsky}. 
This regime will be focused on in the following discussion. 
Then, in the space in which only one density is nonvanishing 
($n_{i} = |\Psi_{i}|^2 \neq 0$ and $n_{j} = 0$, for $i \neq j$), 
the density profile is given by the single-component Thomas-Fermi profile 
$n_{i}=[\mu_{i}-V_{i}({\bf r})]/g_{i}$. Since the Thomas-Fermi approximation 
fails to describe the domain boundary region, more careful analysis is necessary 
to determine the explicit density profile in this region \cite{Barankov,Schaeybroeck}. 

\subsection{Numerical results}\label{numerical}
There are a few papers discussing the properties of immiscible two-component BECs 
under rotation. Kasamatsu {\it et al.} revealed numerically that the winding vortex sheets 
appears as a stable vortex state instead of the periodic vortex lattice \cite{Kasamatsu}. 
Woo {\it et al}. studied Tkachenko excitations in rotating two-component BECs, finding 
that some highly excited Tkachenko modes can give rise to the shear flow of vortices 
along the vortex sheet \cite{Woo}. In a slowly rotating limit, Malomed {\it et al}. studied the 
stability of the rotating cross pattern of domain walls separating one component in the first and 
third quadrant in the $x$-$y$ plane from the other in the second and forth quadrant \cite{Malomed2}. 

Here, we reveal the details of the structure of the vortex sheet 
by numerically solving Eq. (\ref{bingp1}). 
In the following, we consider the two-dimensional problem by assuming the 
translation invariance along the $z$ axis ($\omega_{zi}=0$), 
reducing the original wave function as $\Psi_{i}({\bf r}) 
= \psi_{i}(x,y)/ \sqrt{R_{z}}$ with the typical size $R_{z}$ along 
the $z$-direction. 
It is convenient to measure the length, time and energy scale in units of 
$b_{\rm ho}=\sqrt{\hbar/2 m_{12} \bar{\omega}}$, 
$\bar{\omega}^{-1}$, and $\hbar \bar{\omega}$, respectively, 
with $\bar{\omega} = (\omega_{1} + \omega_{2}) / 2$.
Replacing the wave function as $\psi_{i} \rightarrow \sqrt{N_{i}} \psi_{i}/b_{\rm ho}$, 
we obtain the dimensionless two-dimensional GP equations
\begin{eqnarray}
\biggl( -\frac{m_{12}}{m_i} \nabla^{2} + \frac{m_{i}}{4m_{12}} 
\frac{\omega_{i}^2}{\bar{\omega}^2}  r^{2} 
+ u_{i} |\psi_{i}|^{2} + u_{i,3-i} |\psi_{3-i}|^{2} \nonumber \\ 
- \Omega L_{z} \biggr) \psi_{i} = \mu_{i} \psi_{i}.
\label{nondimgpeq}
\end{eqnarray} 
Here, we define two-dimensional coupling constants 
$u_{i}=8 \pi a_{i} N_{i} (m_{12}/m_i) / R_z$ 
and $u_{i,3-i}=4 \pi a_{12} N_{3-i} / R_z$. 
Since the particle number of each component is conserved, 
the normalization of the wave functions is taken as $\int d {\bf r} |\psi_{i}|^{2} = 1$. 
Using the imaginary time propagation of the time-dependent version of 
Eq. (\ref{nondimgpeq}), we calculate the equilibrium solutions after sufficient 
convergence of some quantities, such as the total energy, of the system.

\begin{figure}
\begin{center}
\includegraphics[width=0.92\linewidth,keepaspectratio]{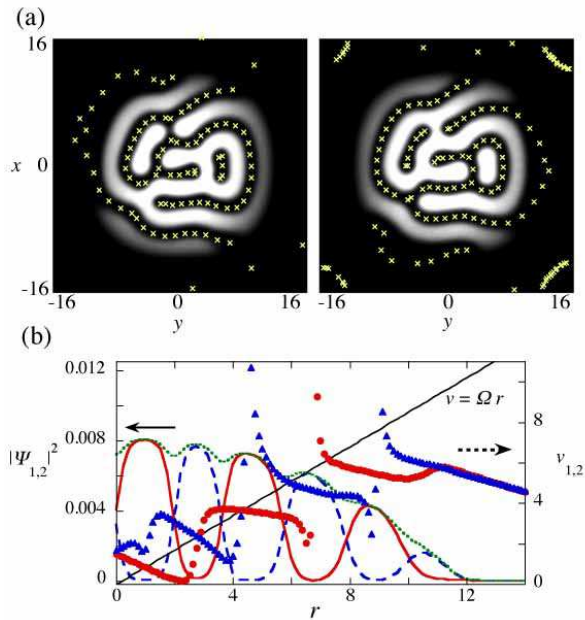}
\end{center}
\caption{(Color online) Typical structure of the vortex sheets, obtained by the numerical 
simulation of the coupled Gross-Pitaevskii equations. The parameter values are 
$u_1 = u_2 = 4000$, $a_{12}/a = \delta=1.1$, and $\Omega=0.85 \omega$ (see the text). 
(a) The contour plots of the two-dimensional density profiles of the two-component 
condensates $|\psi_1|^2$ and $|\psi_2|^2$. 
The vortex sheets are made up of chains of singly quantized vortices whose positions 
are marked by $\times$. (b) Solid, dashed, and dotted curves show the radial density 
profiles of $|\psi_1|^2$, $|\psi_2|^2$, and the total density $n_{\rm T} = 
|\psi_1|^2 + |\psi_2|^2$, respectively. Circular and triangular dots represent the 
corresponding velocity profiles $|{\bf v}_{i}|=\sqrt{(\partial_x \theta_i)^2+(\partial_y \theta_i)^2}$ 
with $i$=1 and 2 (in unit of the length $b_{\rm ho}$ and time $\omega^{-1}$). The velocity of rigid body 
rotation $v =\Omega r$ is shown for comparison.} 
\label{sheetexample}
\end{figure}
First, for simplicity, we assume $m_1 = m_2 =m$, 
$\omega_1 = \omega_2 = \omega$, and $N_1=N_2=N$. 
The typical example of the vortex sheet structure is shown in Fig. \ref{sheetexample}, 
where we used the parameter values as $u_1 = u_2 = u = 4000$, 
$u_{12}/u \equiv \delta = 1.1$, and $\Omega = 0.85 \omega$. 
The two components undergo phase separation to form several density 
domains of the same component. Concurrently, as denoted by crosses in 
Fig. \ref{sheetexample}(a), the nucleated vortices 
merge to form a winding sheet structure like ``serpentine" instead of 
forming a periodic lattice. 
The vortices of the $\psi_1$ component are located 
at the region of the density domains of $\psi_2$ component. 
This can be understood from the fact that the condensate density of one 
component works as a pinning potential for the vortices in the other component.  
By forming vortex sheets, the condensate achieves remarkable phase separation 
compared to a lattice. 

Figure \ref{sheetexample}(b) shows the radial density and velocity profile for 
each component. The domains make a layered structure in the radial direction 
with a certain width. The total density profile 
$n_{\rm T} = n_1 + n_2$ is nearly smooth even in the presence of sharp 
domain boundaries, being given by the Thomas-Fermi profile with $\delta = 1$
\begin{equation}
n_{\rm T} = \sqrt{\frac{2 (1 - \Omega^2)}{\pi u}} 
\left( 1 - \frac{r^2}{R_{\rm TF}^{2}} \right) \label{totalTF}
\end{equation}
with the Thomas-Fermi radius 
\begin{equation}
R_{\rm TF} = \left[ \frac{8u}{\pi(1-\Omega^{2})} \right]^{1/4}. \label{TFrad}
\end{equation} 
We confirm that for $\delta > 1$ the total density is always 
approximated by Eq. (\ref{totalTF}), if there is not asymmetry of the 
parameters for intracomponents such as $u_1 \neq u_2$. 
On the other side, each radial velocity $v_{i}(r)$ is flat between the vortex sheets, 
jumping across the sheet to follow the velocity of rigid body rotation $v=\Omega r$. 
This behavior is quite similar to that in the superfluid helium shown 
in Fig. \ref{sheetsche}.

\begin{figure}
\begin{center}
\includegraphics[width=1.0\linewidth,keepaspectratio]{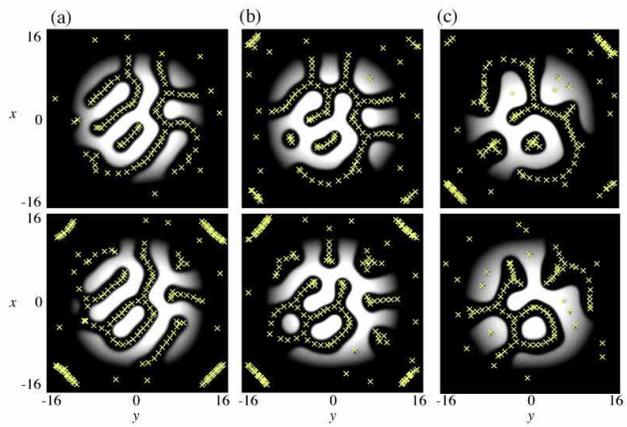}
\end{center}
\caption{(Color online) Typical two-dimensional density profiles and locations of vortices in the 
vortex sheet configuration for $u=4000$ and $\delta=1.5$ (a), $\delta=2.0$ (b), 
$\delta=3.0$ (c). The upper and lower panels show 
$|\psi_1|^2$ and $|\psi_2|^2$, respectively.} 
\label{droplet}
\end{figure}
As $\delta$ increases, the intercomponent repulsion becomes stronger, 
so that the system favors the smaller region of the domain boundary and the larger 
size of the density domain. As a result, the intersheet distance 
gradually increases and the clear serpentine structure disappears  (Fig. \ref{droplet}).
In the limit of large $\delta$, the equilibrium state eventually evolves 
``rotating droplets," as shown in the inset of Fig. \ref{phasedia}, where 
vortex chains do not penetrate into the condensate domains and each 
center of mass of the condensate carries the angular momentum.
This structural change occurs when the equilibrium domain size (sheet distance) 
becomes comparable with the size of the condensate (Thomas-Fermi radius), 
discussed in the next subsection. 

We find that there exist different metastable vortex sheet configurations 
with almost the same energies, which results from the accidental and 
inevitable dislocations nucleated during the imaginary time propagation from different 
trial initial configurations. As one can see below, the energy of the vortex-sheet state 
is mainly determined by two contributions, kinetic energy of the superflow and the 
surface tension, rather than the shape of the vortex sheets \cite{Kasamatsu,Woo}. 
One general feature found here is that most vortex sheets are wound. 
This is in contrast to the consideration by Woo {\it et al}. \cite{Woo}, who expect that 
the most energetically lowered configuration could be the straight vortex sheets 
(shown in the inset of Fig. \ref{phasedia}) instead of the winding sheets. 
The straight sheet configuration is a relatively periodic structure, so that 
this could be obtained near the border of the phase-separation regime $\delta \neq 1$; 
for $\delta < 1$ the stable structure is actually periodic 
vortex lattices \cite{Mueller,Kasamatsu}. For superfluid He$^{3}$ in a rotating 
container, numerical simulations showed that the enertgetically lowest 
configuration is the winding sheets with double spiral pattern, consisting of 
a single piece of sheet with two connections to the wall \cite{Eltsov}. 
 
\begin{figure}
\begin{center}
\includegraphics[width=1.0\linewidth,keepaspectratio]{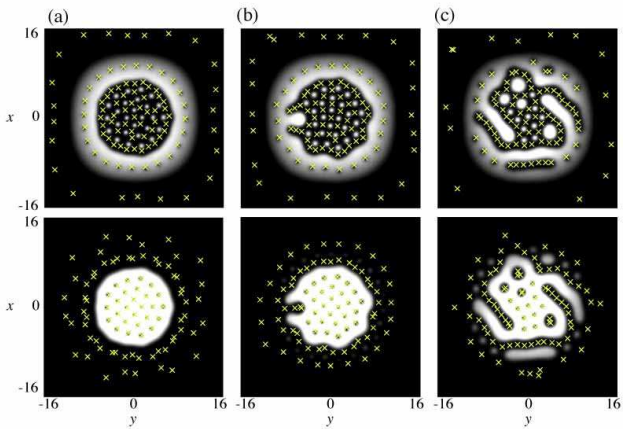}
\end{center}
\caption{(Color online) Typical two-dimensional density profiles and locations of vortices 
in the vortex states for imbalanced intracomponent parameters; $\delta=1.1$, $u_1=4000$, and 
$u_2=3000$ (a), 3500  (b), and 3900  (c). The upper and lower panels show 
$|\psi_1|^2$ and $|\psi_2|^2$, respectively.} 
\label{imbalance}
\end{figure}
When the intracomponents have asymmetric coupling constants $u_1 \neq u_2$, 
which corresponds to the system with imbalance of the populations, 
the intracomponent scattering lengths, or the mass \cite{massdif}, the winding vortex sheet 
cannot be obtained, as shown in Fig. \ref{imbalance}. 
One component with a smaller coupling constant is confined 
at the center of the trap, and the other component surrounds like a 
shell \cite{Ho,Pu,Tim}. Since the condensates contain vortices, this can 
be seen as the rotating core-shell structure. 
In Fig. \ref{imbalance}(a), some density fractions of the $\psi_{1}$ component 
are trapped by the vortex cores of the $\psi_{2}$ component. As a result, the vortex 
lattice in the $\psi_{2}$ component possesses square symmetry due to the 
antiferromagnetic nature of the vortex interactions \cite{Kasamatsu}.
With decreasing the imbalance between $u_1$ and $u_2$, the surface of 
the inner component is distorted to allow the 
penetration of a chain of vortices, as shown in Figs. \ref{imbalance}(b) and \ref{imbalance}(c). 
Therefore one can expect that the clear vortex sheet can be observable for 
the small parameter imbalance between two components. In experiments, 
since tuning of both the intra- and intercomponent scattering lengths is available 
by using the Feshbach resonance \cite{Papp}, such imbalance could be 
reduced under the suitable condition. 

\subsection{Estimation of the intersheet distance} \label{analytical}
\begin{figure}
\begin{center}
\includegraphics[width=0.85\linewidth,keepaspectratio]{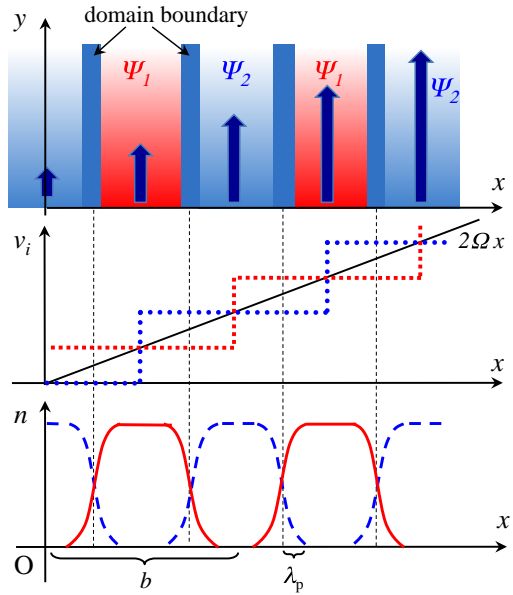}
\end{center}
\caption{(Color online) Simple model to calculate the intersheet distance $b$ for the vortex 
sheet in two-component BECs. The rectangular two-component domains align periodically 
and alternately along the $x$ axis with small domain-boundary region 
characterized by the penetration depth $\lambda_{p}$. The total density is uniform all 
over the space. The condensate velocity ${\bf v}_{i} = v_{i} \hat{\bf y}$ has a 
staircase like increase along the $x$ axis; the value jumps $2 \Omega b$ across 
the sheet for each component and follows the rigid body velocity $2 \Omega x$.  } 
\label{sheetmodel} 
\end{figure}
To understand  the vortex sheet structure more clearly, we give analytical discussions 
by estimating the intersheet distance $b$. 
According to the numerical result in Fig. \ref{sheetexample} and the discussion in Sec. \ref{helium3},
we introduce a simple planar model for two-component BECs as shown in Fig. \ref{sheetmodel}. 
We assume that there is no asymmetry in the intracomponent parameters: $u_1=u_2=u$, 
$m_1=m_2=m$, and $\omega_1=\omega_2=\omega$. 
Here, the two components form straight layered vortex sheets along the 
$y$ axis, the sheet distance $b$ being defined as the length between one sheet 
and its nearest sheet in the {\it same} component. 
This corresponds to the view of the cylindrical planar sheets without defects 
in the polar coordinates instead of the Cartesian ones. In equilibrium, the condensate 
exhibits the rigid body rotation $\nabla \times {\bf v}_{\rm rb} = 2 \Omega \hat{\bf z}$, 
following the rotation of the reference frame of the system; 
we take ${\bf v}_{\rm rb} = 2 \Omega x \hat{\bf y}$.  
However, the actual velocity of the condensate has a staircase like change 
along the $x$ axis because of the vortex sheets. 
In our model, the velocity of each component is assumed to be 
uniform between the vortex sheets of the {\it same} component; 
the value of $v_{i}$ increases by $2 \Omega b$ across 
every sheet. We neglect the effect of the trapping potential, so that the 
total density $n_{\rm T}=n_{1}+n_{2}$ is approximately 
uniform even in the presence of the domain boundaries with the penetration 
depth $\lambda_{\rm p}$ of the order of the healing length $\xi=\hbar/\sqrt{2m\mu}$ 
\cite{Ao,Barankov,Schaeybroeck}. 
Although the actual condensate density is nonuniform because of the 
trapping potential, this model could be valid for condensates with large size 
in the central region of the trapping potential.

We calculate the energy $E$ of Eq. (\ref{energyfunctio2}) per volume 
in the range $0<x<b$ with the sheet distance $b$ being determined 
so as to minimize $E$ per unit area. The $b$-dependent energy 
comes from the kinetic energy of the condensate flow and the excess energy 
due to the domain boundaries, namely the surface 
tension $\sigma$ \cite{Ao}. The flow energy per volume in the rotating frame is 
\begin{eqnarray}
\frac{1}{b} \int_{0}^{b} dx \sum_{i} m_{i} n_{i} \frac{(v_{i}-2 \Omega x)^{2}}{2}  
= \frac{1}{48} m n_{\rm T} \Omega^{2} b^{2}. 
\end{eqnarray}
Here, the integral was done within the range $[b/4,3b/4]$ ($[0,b/4]$ and $[3b/4,b]$) 
for $i=1$ (2) by neglecting the penetration of the domain. 
Because of the constant density, the interaction energy is independent of the 
sheet distance $b$. Then, the $b$-dependent energy per unit volume is written by 
\begin{equation}
\frac{E}{b} =\frac{1}{48} m n_{\rm T} \Omega^{2} b^{2} + \frac{2\sigma}{b}, 
\end{equation}
where we included the surface tension from two domain boundaries within $0<x<b$. 
Minimizing this energy with respect to $b$, one obtains 
\begin{equation}
b=2 \left( \frac{6 \sigma}{m n_{\rm T} \Omega^{2}} \right)^{1/3}.  \label{sheetd}
\end{equation}
The analytical expression of the surface tension $\sigma$ can be found 
in Refs. \cite{Barankov,Schaeybroeck}; 
for $m_1=m_2=m$ and $a_1=a_2=a$, the surface tension in a weak 
segregated regime ($\delta \simeq 1$) is given by 
\begin{equation}
\sigma_{\rm w} = 2 P \xi \sqrt{\delta-1},  \label{weaktension}
\end{equation}
and that in a strongly segregated regime  ($\delta > 1$) is given by 
\begin{equation}
\sigma_{\rm s} = 4 P \xi \left[ \frac{2\sqrt{2}}{3} - \frac{0.514}{\delta^{1/4}} 
- 2\left(\frac{0.055}{\delta^{3/4}} + \frac{0.067}{\delta^{5/4}} \right) \right] 
\label{strongtension}
\end{equation}
with the healing length $\xi = \hbar/\sqrt{2 m \mu}$ and the equilibrium pressure 
$P = \mu_1^{2} / 2 g_1 = \mu_2^{2} / 2 g_2$. 
These two regimes have a crossover near $\delta  \simeq 1.34$, 
according to Fig. 3 in Ref. \cite{Schaeybroeck}.
In the following, we use Eq. (\ref{weaktension}) for $\delta < 1.34$ and  
Eq. (\ref{strongtension}) for $\delta > 1.34$ to evaluate Eq. (\ref{sheetd}). 

\begin{figure}
\begin{center}
\includegraphics[width=0.9\linewidth,keepaspectratio]{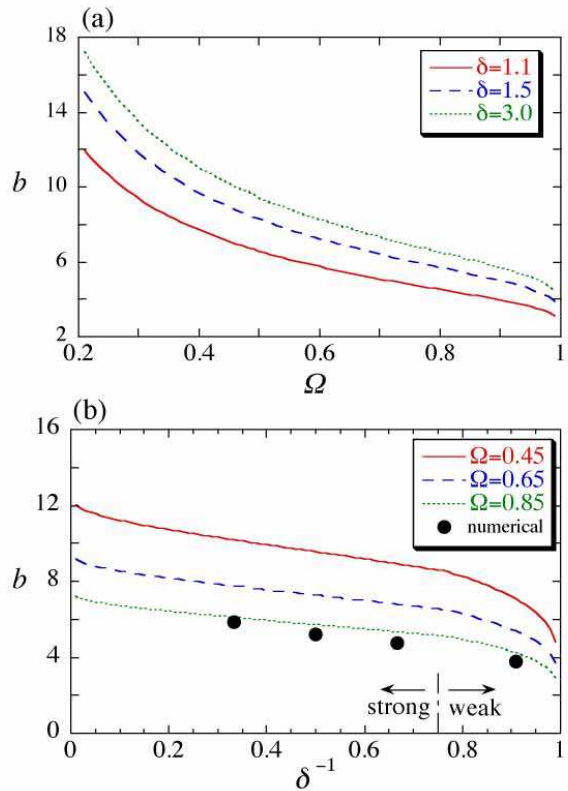}
\end{center}
\caption{(Color online) The sheet distance $b$ (in unit of $b_{\rm ho}$) 
as a function of $\Omega$ (a) and $\delta^{-1}$ (b) for $u=4000$. 
In (b), the two results using Eqs. (\ref{weaktension}) and (\ref{strongtension}) 
are smoothly connected at $\delta^{-1} = 0.75$. Also, the numerically estimated 
values of $b$ for $\Omega = 0.85$ are plotted for $\delta =$ 1.1 
(Fig. \ref{sheetexample}), 1.5, 2.0, and 3.0 (Fig. \ref{droplet}); 
the values were taken by averaging the intersheet distance 
from the profile of vortex positions.} 
\label{sheetdist}
\end{figure}
By using the Thomas-Fermi density with $\delta=1$ [Eq. (\ref{totalTF})] at $r=0$ 
as the value of $n_{\rm T}$ in the denominator of Eq. (\ref{sheetd}), 
one can estimate the approximate sheet distance $b$ for the trapped two-component BECs.
Figure \ref{sheetdist} represents the value of $b$ as a function of $\Omega$ 
and $\delta^{-1}$. The two results using Eqs. (\ref{weaktension}) 
and (\ref{strongtension}) are smoothly connected at $\delta^{-1} = 0.75$. 
The sheet distance monotonically decreases (increases) as the rotation (the intercomponent 
repulsion) becomes faster (stronger). 
We compare our numerically estimated values of $b$ taken from 
Figs. \ref{sheetexample} and \ref{droplet}, 
obtaining excellent agreement with Eq. (\ref{sheetd}).

\begin{figure}
\begin{center}
\includegraphics[width=0.9\linewidth,keepaspectratio]{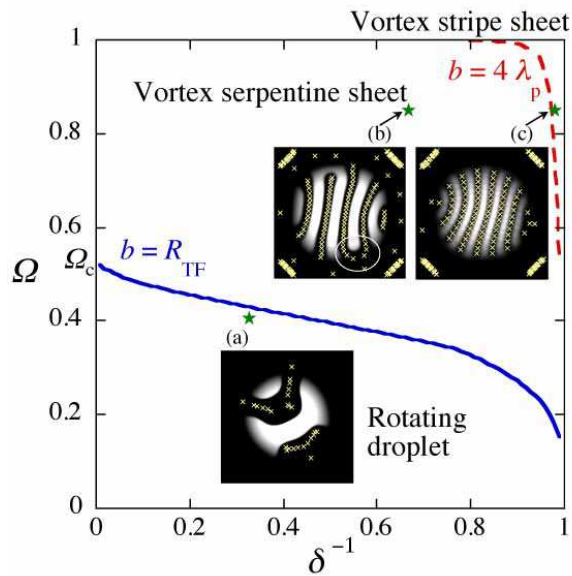}
\end{center}
\caption{(Color online) $\Omega$-$\delta^{-1}$ phase diagram of the 
equilibrium vortex states. The borders are determined by comparing the 
characteristic length scales (see the text). The insets show the 
typical equilibrium structures of the density $|\psi_1|^2$ and 
the vortex positions for the parameters indicated by the star marks; 
(a) $(\Omega,\delta) = (0.4, 3.0)$, (b) $(\Omega,\delta) = (0.85, 1.5)$, 
(a) $(\Omega,\delta) = (0.85, 1.02)$. In (b), the region indicated by a circle 
represents the linked edges of two vortex sheets. } 
\label{phasedia}
\end{figure}
Besides the sheet distance, we have the other characteristic length scales 
in this system: (i) the penetration depth $\lambda_{\rm p}$ of the domain 
boundary; for the weakly segregated regime, this is given by 
$\sim \xi/\sqrt{\delta-1}$ \cite{Schaeybroeck}, (ii) the Thomas-Fermi radius 
$R_{\rm TF}$ given by Eq. (\ref{TFrad}). 
By comparing these lengths with $b$, we can identify three distinct structural phases, 
summarized by the phase diagram in Fig. \ref{phasedia}.
The vortex sheet is expected in the region $b<R_{\rm TF}$. 
The border $b=R_{\rm TF}$ is gradually increased as $\delta$ goes to infinity, where 
we can get a critical rotation frequency 
$\Omega_c = [16 \pi/(\sqrt{\pi u} + 16 \pi)]^{1/2}$ above which the vortex sheet 
is observable for arbitary $\delta$. 
In the region $b > R_{\rm TF}$, the vortex sheet cannot penetrate 
into the condensate domain, the condensates forming droplets with a center-of-mass 
rotation [see the inset (a) of Fig. \ref{phasedia}]. Since most of the vortices are located 
in the low-density region as ``ghosts" \cite{Tsubota}, they contribute little to 
the energy cost. Thus, the rotating droplet can exist even for very small $\Omega$. 

In the region $b<R_{\rm TF}$, the equilibrium structure is generally 
the serpentine vortex sheets. However, when $b$ becomes comparable to 
the penetration depth $\lambda_{\rm p}$, the periodic stripe structure 
of vortex sheets is formed [Fig. \ref{phasedia}(c)] instead of the serpentine one. 
We determine this border at $b=4\lambda_{\rm p}$, 
where the density plateau for each component is 
vanished in our model of Fig. \ref{sheetmodel}. 
In this weakly segregated regime, the influence of the periodicity of vortex states 
for $\delta \leq 1$ survives. As $b$ increases with $\delta$, 
the domain size increases, which leads to the unfavorable tight confining of the 
relevant number of vortices into the decreasing low-density regions. 
As a result, the stripe configuration is disrupted by linking edges of the neighboring sheets; 
the inset (b) in Fig. \ref{phasedia} is obtained by the imaginary time propagation 
and the gradual increase in $\delta$ from the state (c). Therefore a tendency to confine
vortices as loosely as possible and the inevitable dislocations give rise to the 
serpentine vortex sheets such as Figs. \ref{sheetexample} and \ref{droplet}, although 
the total energy of the quasistripe state (c) is lower than those configurations. 
This mechanism is involved when the vortex sheets is created experimentally 
by rotating potentials.

\section{CONCLUSION} \label{concle}
We have revealed the equilibrium structures of the rotating 
two-component BECs in a phase-separated regime. If the imbalance of 
the intracomponent parameter is sufficiently small, the two-component BECs form a 
winding vortex sheet structure; otherwise, the condensates form 
a rotating core-shell structure. 
We developed a simple model to estimate the vortex sheet structure, 
following the argument of superfluid helium by Landau and Lifshitz. 
As the intercomponent coupling becomes larger, 
the sheet distance is also increased, approaching the radius of 
the condensate density. With this increase, the sheet structure changes 
from the stripe pattern to the winding one like ``serpentine." In the 
slowly rotating case, the sheet distance exceeds the condensate radius so that 
the condensate forms rotating droplets, where the angular momentum
was carried by their center-of-mass rotation. 
These results are summarized in the phase diagram of Fig. \ref{phasedia}, obtained 
by the comparison of the characteristic length scales, the sheet distance $b$, the 
penetration depth $\lambda_{\rm p}$, and the Thomas-Fermi radius $R_{\rm TF}$. 

Two-component BECs with tunable intercomponent interactions were 
obtained experimentally \cite{Thalhammer,Papp}. 
By rotating them, it is possible to explore the 
interesting regime of vortex phases in multicomponent BECs.  

\begin{acknowledgments}
K.K. acknowledges the supports of a Grant-in-Aid for Scientific Research from JSPS (Grant No. 18740213). 
M.T. acknowledges the supports of a Grant-in-Aid for Scientific Research from JSPS (Grant No. 18340109) 
and a Grant-in-Aid for Scientific Research on Priority Areas from MEXT (Grant No. 17071008). 
\end{acknowledgments}


\end{document}